\documentclass[aps,prr,twocolumn,superscriptaddress,showpacs,preprintnumbers,amsmath,amssymb,longbibliography]{revtex4-2}
\usepackage{graphicx}% Include figure files
\usepackage{dcolumn}% Alig table columns on decimal point
\usepackage{bm}% bold math
\usepackage{enumerate}
\usepackage{mathrsfs}
\usepackage{amsmath}
\usepackage{multirow}
\usepackage{subfigure}
\usepackage{braket}
\usepackage{epstopdf}
\usepackage[titletoc]{appendix}
\usepackage[colorlinks=true,breaklinks=true,linkcolor=blue,citecolor=blue,urlcolor=blue]{hyperref}
\usepackage{color}
\usepackage{fmtcount}
\usepackage{multirow}

\begin{document}

\def\rtx@apsprl{\class@info{APS journal PRL selected}}
	
\title{Noise-resistant off-resonance $\pi$-rotation realized in an NMR system
}	
%\title{Experimental realization of noise-resistant off-resonance $\pi$-rotation in a NMR system}	
%\title{Physical spectral decomposition of weakly coupled nuclear spins by quantum state preparation}
	
\author{Xilong Wang}
\altaffiliation{These authors contributed equally to this work.}
\affiliation{School of Physics and Technology, Wuhan University, Wuhan, Hubei 430072, China}
\affiliation{School of Physics, Hangzhou Normal University, Hangzhou, Zhejiang 311121, China}
\affiliation{Hefei National Laboratory, Hefei, Anhui 230088, China}

\author{Peng Du}
\altaffiliation{These authors contributed equally to this work.}
\affiliation{School of Physics and Technology, Wuhan University, Wuhan, Hubei 430072, China}

\author{Tianzi Wang}
\affiliation{School of Physics and Technology, Wuhan University, Wuhan, Hubei 430072, China}
\affiliation{School of Physics, Hangzhou Normal University, Hangzhou, Zhejiang 311121, China}
\affiliation{Hefei National Laboratory, Hefei, Anhui 230088, China}

\author{Haoxue Qiao}
\email[Corresponding email: ]{qhx@whu.edu.cn}
\affiliation{School of Physics and Technology, Wuhan University, Wuhan, Hubei 430072, China}

\author{Wenxian Zhang}
\email[Corresponding email: ]{wxzhang@hznu.edu.cn}
\affiliation{School of Physics, Hangzhou Normal University, Hangzhou, Zhejiang 311121, China}

\begin{abstract}
Precise manipulation of qubits (spins) requires strong suppression of noise arising from couplings between qubits and their environment, as well as from driving field fluctuations. We propose a double-pulse sequence and an eighteen-pulse sequence to realize a single spin $\pi$-rotation under off-resonance conditions (instead of conventional on-resonance or near-resonance conditions). It is proved that the latter sequence performs much better than the former in simultaneously suppressing both dephasing and driving field noise. As a proof of principle, we experimentally demonstrate, by implementing both sequences in an NMR spin system with a polarized initial spin state, that the off-resonance $\pi$-rotation fidelity reaches as high as $0.9992\pm 0.0008$ for the double-pulse sequence and $0.9997\pm 0.0006$ for the eighteen-pulse sequence. In addition, the noise-tolerance range of the eighteen-pulse sequence is about 20\% of the driving field strength, approximately three times wider than that of the double-pulse sequence. Besides its potential application in the NMR community, our proposal provides an alternative practical way to accurately control qubits by suppressing the coexisting transversal and longitudinal noises.
\end{abstract}

\date{\today}

\maketitle

%%%%%%%%%%%%%%%%%%%%%%%%%%%%%%%%%%%%%%%%%%%%%%%%%%%%%%%%%%%%%%%%%%%%%%%%%%%%%%%
\section{\label{sec:Introduction}Introduction}
Quantum computing has developed rapidly and demonstrated great potential compared to classical computing in machine learning~\cite{PhysRevResearch.7.013007,PRXQuantum.4.010328}, pharmaceutical research~\cite{article2,Mensa2023}, communication security~\cite{Liao:2017isj}, artificial intelligence~\cite{article,Abbas:2020qrn,doi:10.1126/science.abn7293}, and so on. Although a logic qubit was experimentally realized very recently~\cite{Egan:2021xzp,Zhang2025}, most advanced research in quantum computing today remains in the noisy intermediate-scale quantum (NISQ) era~\cite{Preskill2018quantumcomputingin,md6x-pfrr}.  Precise control of qubit gates is still the core requirement. However, environmental noise can severely undermine the operational accuracy of quantum gates~\cite{PhysRevLett.82.2417,doi:10.1126/science.279.5349.342}. To suppress noise, many quantum technologies have been developed, including quantum error correction, decoherence free subspace, dynamically corrected gates~\cite{PhysRevLett.81.2152,Lidar_Brun_2013,PhysRevLett.102.080501,RevModPhys.87.307,PhysRevA.109.062411}. Among them, dynamically corrected gates with composite pulse sequences are often employed due to their high efficiency in noise suppression as well as thier simplicity in experimental implementation~\cite{PhysRevB.96.195424,PhysRevA.98.012301,PhysRevA.90.012316}. 
 
In experiments, noise mainly comes from two orthogonal directions~\cite{Nielsen_Chuang_2010,article4,doi:10.1126/science.279.5349.342,PhysRevLett.106.240501,PhysRevLett.105.200402,PhysRevLett.113.220501}: longitudinal noise originating from the non-uniform distribution of the main static magnetic field used to eliminate the degeneracy of the two-level system, and transversal noise introduced by the power fluctuation of the control field. Composite pulse control, originating from the field of nuclear magnetic resonance (NMR), is cleverly designed to suppress noise by adjusting the timing, amplitude and phase of multiple basic pulses~\cite{WIMPERIS1994221,RevModPhys.76.1037,LEVITT198661}. Previously, optimal pulse control technology has been widely applied in many experimental platforms to improve the precision of qubit gates, such as nitrogen-vacancy centers~\cite{articleDobrovitski,article4}, superconducting qubits~\cite{d9yg-d3qr,PhysRevLett.93.157005}, and quantum-dots~\cite{Yoneda:2017fdy}. To facilitate operation, most composite pulses require the system and the control field to be on-resonance or near-resonance. These include the BB1 sequence~\cite{WIMPERIS1994221}, which suppresses transversal pulse errors, and sequences that suppress longitudinal noise, such as the CORPSE sequence~\cite{PhysRevA.67.042308} and the composite pulse sequence proposed by Brown {\it et al}.~\cite{PhysRevA.70.052318, ALWAY2007114}. However, for certain quantum systems, due to the encoding characteristics of the qubits, it is inevitable to perform control on a single qubit under off-resonance condition~\cite{articleWang,articleFoletti,Jirovec:2021rcm}. Furthermore, with the development of large-scale quantum computing, the increasingly stronger control field often causes non-resonant excitation of adjacent qubits (crosstalk), which occurs in superconducting qubits~\cite{articleQuantum,PRXQuantum.3.020301,articleNature,PhysRevLett.129.060501}, trapped ions~\cite{Egan:2021xzp}, and NMR systems~\cite{RevModPhys.76.1037,FREEMAN199859}.

In this work, focusing on the off-resonance conditions, we propose a double-pulse sequence (DP sequence) to achieve a $\pi$-rotation of a spin (qubit). To simultaneously suppress the longitudinal noise caused by stray magnetic fields and the transversal noise from RF pulse power fluctuations, we further develop an optimized composite pulse sequence, namely the noise-resistant eighteen-pulse sequence (NREP sequence) which nests the DP sequence into the BB1 sequence~\cite{WIMPERIS1994221} and the composite pulse proposed by Brown {\it et al}.~\cite{PhysRevA.70.052318, ALWAY2007114}. As a proof of principle, we experimentally demonstrate the performance of the DP and NREP sequences in noise suppression in a liquid NMR system consisting of $C_2 F_3 I$ molecules. Our proposal points to a new direction for realizing high-precision $\pi$-rotation under the off-resonance conditions and provides a possible way to solve the crosstalk problem caused by hard pulses in two-qubit systems. 

\section{\label{sec:model}Noise-resistant off-resonance $\pi$-rotation}
We consider the evolution of a single spin-$1/2$ in a strong bias magnetic field $B_0$ along the $z$ axis and an RF pulse with a carrier frequency $\omega$ and initial phase $\phi$ in the $xy$ plane. The Hamiltonian (setting $\hbar=1$) is
$H=(\omega_0/ 2)\sigma_z+(\Omega/2) [\cos(\phi+\omega t)\sigma_x +\sin(\phi+\omega t)\sigma_y]$, where $\vec{\sigma}=(\sigma_{x},\sigma_{y},\sigma_{z})$ represents the Pauli operators, $\omega_0 \equiv \gamma B_0$ with $\gamma$ being the gyromagnetic ratio of the spin, and the RF pulse Rabi frequency $\Omega\equiv\gamma B_1$ with $B_1$ being the effective transversal magnetic field. In the interaction picture with $H_0=(\omega/2) \sigma_z$, the above Hamiltonian is simplified as
\begin{equation}
	H = {\Delta\over 2}\sigma_z + {\Omega\over 2}(\sigma_x \cos \phi+\sigma_y \sin \phi),
	\label{eq2}
\end{equation}
where the detuning $\Delta \equiv \gamma \Delta B_0 = \omega_0-\omega$. 

Under the off-resonance condition $\Delta = \Omega$, which benefits us in pulse design and noise suppression, it is easy to prove that a $\pi$-rotation of the spin can be realized by applying a DP sequence (see Sec. $\mathrm{\uppercase\expandafter{\romannumeral 1}}$ of the Supplemental Materials~\cite{sm}). This DP sequence includes two $\pi$-pulses, $(\pi)_{\phi'} - (\pi)_{\pi+\phi'}$, where $\phi'$ and $\pi + \phi'$ denote respectively the phase of the first and the second RF pulse as in Eq.~(\ref{eq2}). One finds immediately that the direction $\vec B_e$ of the first pulse is perpendicular to that of the second one. It is easy to check that the total spin evolution under the DP sequence is $U = \exp(-i\pi\hat n \cdot \vec \sigma/2)$, which is nothing but a $\pi$-roation along the direction $\hat n = [\cos(\phi'+\pi/2), \sin(\phi'+\pi/2), 0]$. 

\begin{figure}[tb]
	\centering
	\includegraphics[width=3.2in]{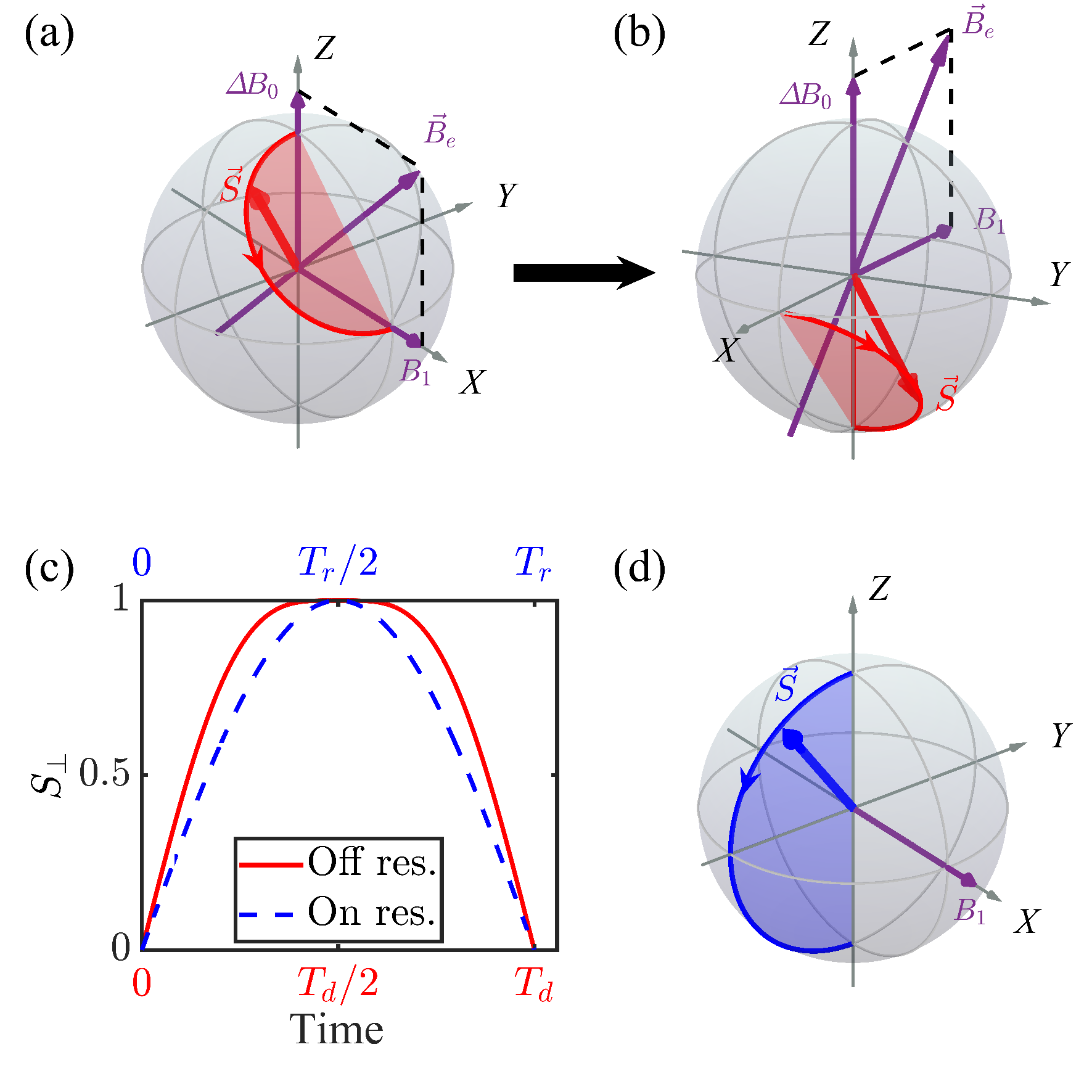}
	\caption{Trajectory of a spin $\vec{S}$ initially along the $+z$ direction under the off-resonance DP sequence: (a) the first pulse with $B_1$ along the $+x$ direction and (b) the second pulse with $B_1$ along the $-x$ direction. The effective rotation axis $\vec{B}_{e}$ is determined by $\Delta B_0$ and $B_1$. (c) Evolution of $S_{\perp}$ under the off-resonance DP sequence (red solid line) and the on-resonance single pulse (blue dashed line). The total evolution period is $T_d = \sqrt 2 \pi / \Omega$ for the DP sequence and $T_r = \pi / \Omega$ for the resonant pulse. (d) Trajectory of a spin $\vec{S}$ with an initial direction along the $+z$ axis under the resonant pulse with $\phi' = 0$. The rotation axis is along the $+x$ direction.}
	\label{f1}
\end{figure}

To visualize the spin trajectory, we employ the spin average $\vec{S}=(S_x, S_y, S_z)  \equiv  (\langle\sigma_{x}\rangle, \langle\sigma_{y}\rangle,\langle\sigma_{z}\rangle)$. By setting the initial spin along the $+z$ direction, the process of the $\pi$-rotation under the DP sequence with $\phi'=0$ is shown in Fig.~\ref{f1}(a) and (b). In total, the equivalent rotation axis is along the $+y$ direction.

We compare the performance of the off-resonance DP sequence with the conventional on-resonance single driving pulse to realize a $\pi$-rotation in Fig.~\ref{f1}(c), where $S_\perp = \sqrt{S_x^2 + S_y^2}$. The spin trajectory for the on-resonance case is shown in Fig.~\ref{f1}(d). Clearly, the total evolution time of the $\pi$-rotation under the DP sequence is $\sqrt 2$ times that under the resonant pulse.

In a real experimental environment, the spin may experience many kinds of noises, such as noises from longitudinal and transversal stray magnetic fields and from power fluctuation of the driving field~\cite{Nielsen_Chuang_2010,article4,doi:10.1126/science.279.5349.342,PhysRevLett.106.240501,PhysRevLett.105.200402,PhysRevLett.113.220501}. Due to the strong bias field $B_0$, the transversal magnetic field noises are significantly suppressed in the rotating reference frame and can be safely neglected. The Hamiltonian of the spin in a noisy environment is thus described by
\begin{equation}
	H_n={\Delta\over 2}(1+\beta)\sigma_z+{\Omega\over 2}(1+\epsilon)[\sigma_x \cos(\phi)+\sigma_y \sin(\phi)],
\end{equation}
where $\beta = \delta_0 / \Delta$ with $\delta_0$ characterizing the longitudinal magnetic noise strength and $\epsilon = \delta_1 / \Omega$ with  $\delta_1$ characterizing the driving field noise from the power fluctuation. We assume that $\delta_0$ and $\delta_1$ (in general $\delta_0  \ll \Delta$ and $\delta_1  \ll \Omega$) are quasi-static noise, due to the timescale of the spin rotation being much less than $\delta_0^{-1}$ and $\delta_1^{-1}$. For instance, $\delta_0 / \Delta \sim 0.1\%$ and $\delta_1 / \Omega \sim 2\%$  with $\Delta \approx \Omega \sim 20$~KHz in our NMR experiments.

We adopt the final state fidelity $F$ to evaluate the performance of a composite pulse against the noises $\delta_0$ and $\delta_1$. For a single spin $\pi$-rotation, $F$ is defined as
\begin{equation}
	\begin{aligned}
		F=|\langle{\chi_f}|{\chi_e}\rangle|^2=|\langle{\chi_f}|U_e|{\chi_i}\rangle|^2,
	\end{aligned}
\end{equation}
where the initial state is $|{\chi_i}\rangle=\cos({\theta\over2})|{0}\rangle+e^{i\varphi}\sin({\theta\over2})|{1}\rangle$ and the ideal target state after a perfect $\pi$-rotation is $|{\chi_f}\rangle=-e^{i\varphi}\sin({\theta\over2})|{0}\rangle+\cos({\theta\over2})|{1}\rangle$. Here, the polar angle $\theta \in [0,\pi]$ and the azimuthal angle $\varphi \in [0,2\pi]$ characterize the initial spin direction on the Bloch sphere. The evolved state under the pulses and noises is $|{\chi_e}\rangle = U_e|{\chi_i}\rangle$ with $U_e$ being the evolution operator.

\begin{figure}[tb]
	\centering
	\includegraphics[width=3.25in]{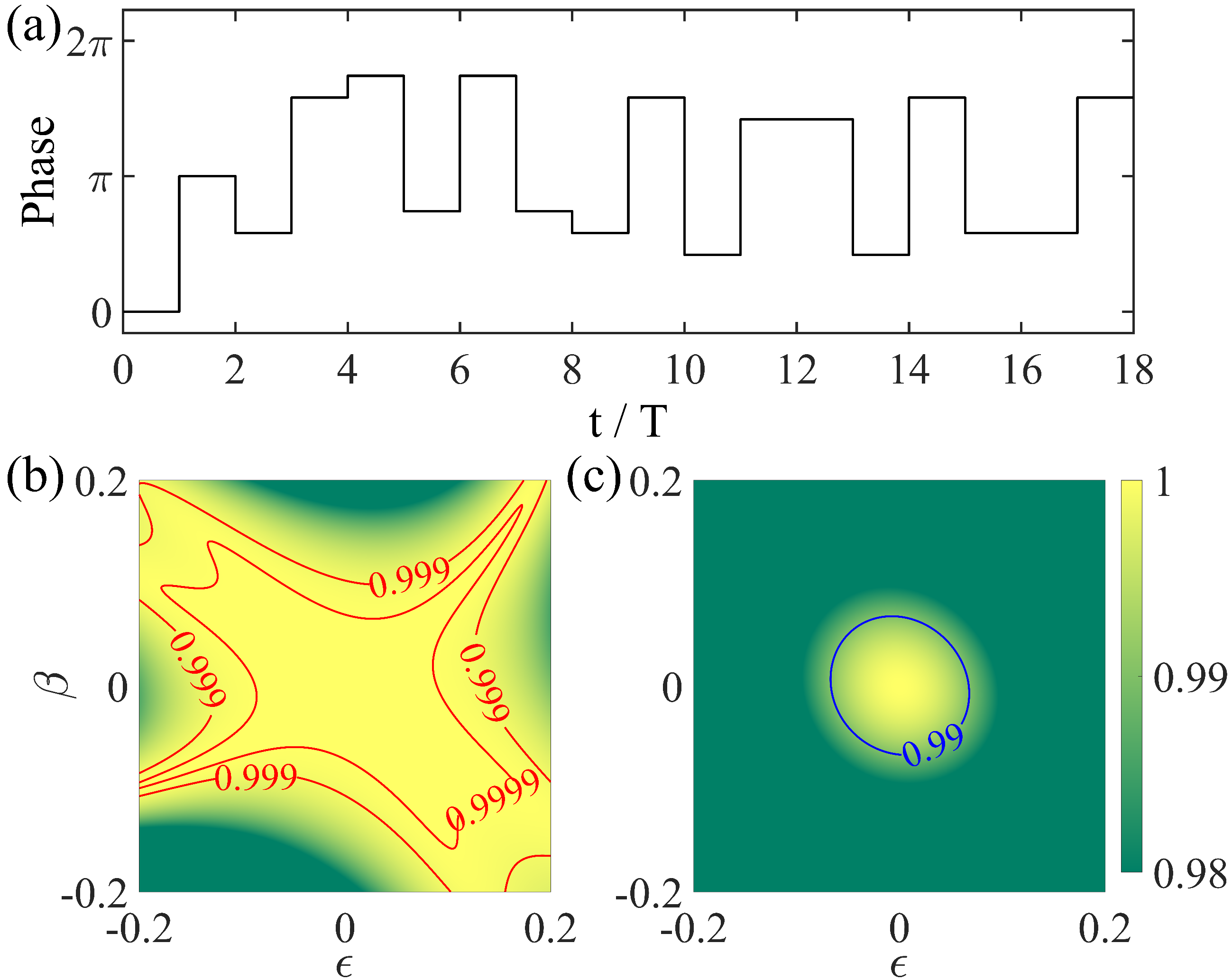}
	\caption{(a) Phase of the NREP sequence with the pulse length $T = \pi /(\sqrt{2} \Omega)$. Final state fidelity $F$ with the initial state along the $+z$-direction after (b) the NREP sequence and (c) the DP sequence. The high fidelity region in (b) is much larger than that in (c), manifesting better noise-suppression of $\delta_0$ and $\delta_1$ for the NREP sequence.}
	\label{f2}
\end{figure}

It is known that the na\"ive on-resonance single pulse and the off-resonance DP sequence $\pi$-rotation are prone to various noises. For example, the fidelity $F_{DP}$ of the DP sequence depends on $\epsilon$ and $\beta$ in a combined quadratic form, resulting in errors of the actual $\pi$-rotation~\cite{sm}. To suppress the noise $\delta_0$ under the near resonance condition, Brown {\it et al}. proposed a composite pulse, $(\pi)_{0^\circ}-(\pi)_{\pi-\phi_1}-(\pi)_{-\phi_1}-(\pi)_{\pi+\phi_1}-(\pi)_{\phi_1}$, where $\pi_\phi$ represents a $\pi$-rotation of the spin about the axis ($\cos\phi, \sin\phi, 0$) on the Bloch sphere and $\phi_1=\arccos(-1/4)$~\cite{PhysRevA.70.052318,ALWAY2007114}.  It is proved that the final state fidelity $F_{1}$ depends on $\beta^4$ to the leading order, under the condition $\epsilon = 0$. Additionally, the BB1 pulse sequence was proposed to eliminate the RF pulse-length error and has been widely applied in spin-1/2 experiments~\cite{WIMPERIS1994221}. This sequence is $(\pi)_{0^\circ}-(\pi)_{\phi_1}-(\pi)_{\phi_2}-(\pi)_{\phi_2}-(\pi)_{\phi_1}$ with $\phi_1=\arccos(-1/4)$ and $\phi_2=3\phi_1 $. Its final state fidelity $F_{BB1}$ depends on $\epsilon^6$ to the leading order, under the condition $\beta = 0$ (See Sec. $\mathrm{\uppercase\expandafter{\romannumeral 2}}$ of the Supplemental Materials~\cite{sm}). However, with both $\beta$ and $\epsilon$ present simultaneously, the above two pulse sequences perform poorly.

We propose an NREP sequence to achieve a precise off-resonance $\pi$-rotation, by nesting the DP sequence into the combination of the BB1 sequence and the one proposed by Brown {\it et al.}. This NREP sequence consists of eighteen pulses and demonstrates excellent robustness to suppress the noises $\delta_0$ and $\delta_1$ simultaneously. The overall rotation is divided into nine DP operations 
$U_e=U_9(\pi,\phi_1)U_8(\pi,\pi+\phi_1)U_7(\pi,-\phi_1)U_6(\pi,\pi-\phi_1)U_5(\pi,\phi_1)U_4(\pi,\phi_2)U_3(\pi,\phi_2) U_2(\pi,\phi_1) U_1(\pi,0)$, where the DP operation $U_k(\pi,\phi)=V_k(\pi,\pi+\phi)V_k(\pi,\phi)$ with $V_k(\pi,\phi) = \exp(-iH_{n}T)$ ($k=1, 2, \cdots, 9$), $T=\pi/(\sqrt{2}\Omega)$, $\phi_1 = \arccos(-1/4)$, and $\phi_2 = 3\phi_1 $. Each DP operation $U_k(\pi,\phi)$ realizes an off-resonance $\pi$-rotation. The RF pulse phase $\phi$ depends piecewise-constantly on time as shown in Fig.~\ref{f2}(a). We set the first pulse phase $\phi = 0$. In the NREP sequence, the first to the fifth DP operations originate from the BB1 sequence and the sixth to the ninth ones from the composite pulse sequence proposed by Brown {\it et al.}~\cite{WIMPERIS1994221, PhysRevA.70.052318, ALWAY2007114}. After a straightforward analysis, we find that the finial state fidelity after the NREP sequence $F_{NREP}$ shows a quartic dependence on the combinations of $\epsilon$ and $\beta$ to the leading order~\cite{sm}, indicating significantly improved robustness against both $\delta_0$ and $\delta_1$, compared to the na\"ive DP sequence. 

To illustrate the performance difference of the NREP and the DP sequence to in noise-suppression, we plot their fidelity over three representative initial spin states, i.e., the initial spin is along $+x$, $+y$, and $+z$ direction. For the spin initially along the $+z$ axis, which is often the starting point (the thermal equilibrium state) in an NMR experiment, Fig.~\ref{f2}(b) and (c) show the fidelity after the NREP and the DP sequence, respectively. For most of the parameter range $\left\{\beta,\epsilon\right\}\in{[-0.2,0.2]}$, the fidelity of the NREP sequence is above $0.999$, manifesting its superior performance in the suppression of noises. On the contrary, to ensure a final state fidelity exceeding 0.99 under the DP sequence, the relative noise strengths $\epsilon$ and $\beta$ must remain below approximately 0.07. Clearly, the NREP sequence exhibits much better noise-suppression than the DP sequence. For other initial spin states lying in the $xy$ plane (see Supplemental Materials~\cite{sm}), the NREP sequence also shows substantially better noise-suppression than the DP sequence. Furthermore, while the performance of the DP sequence exhibits a profound dependence on the azimuthal angle of the initial spin direction, the NREP sequence remains robust across all initial states in the $xy$ plane, demonstrating more uniform noise resistance.

\begin{figure}[tbhp]
	\centering
	\includegraphics[width=3.25in]{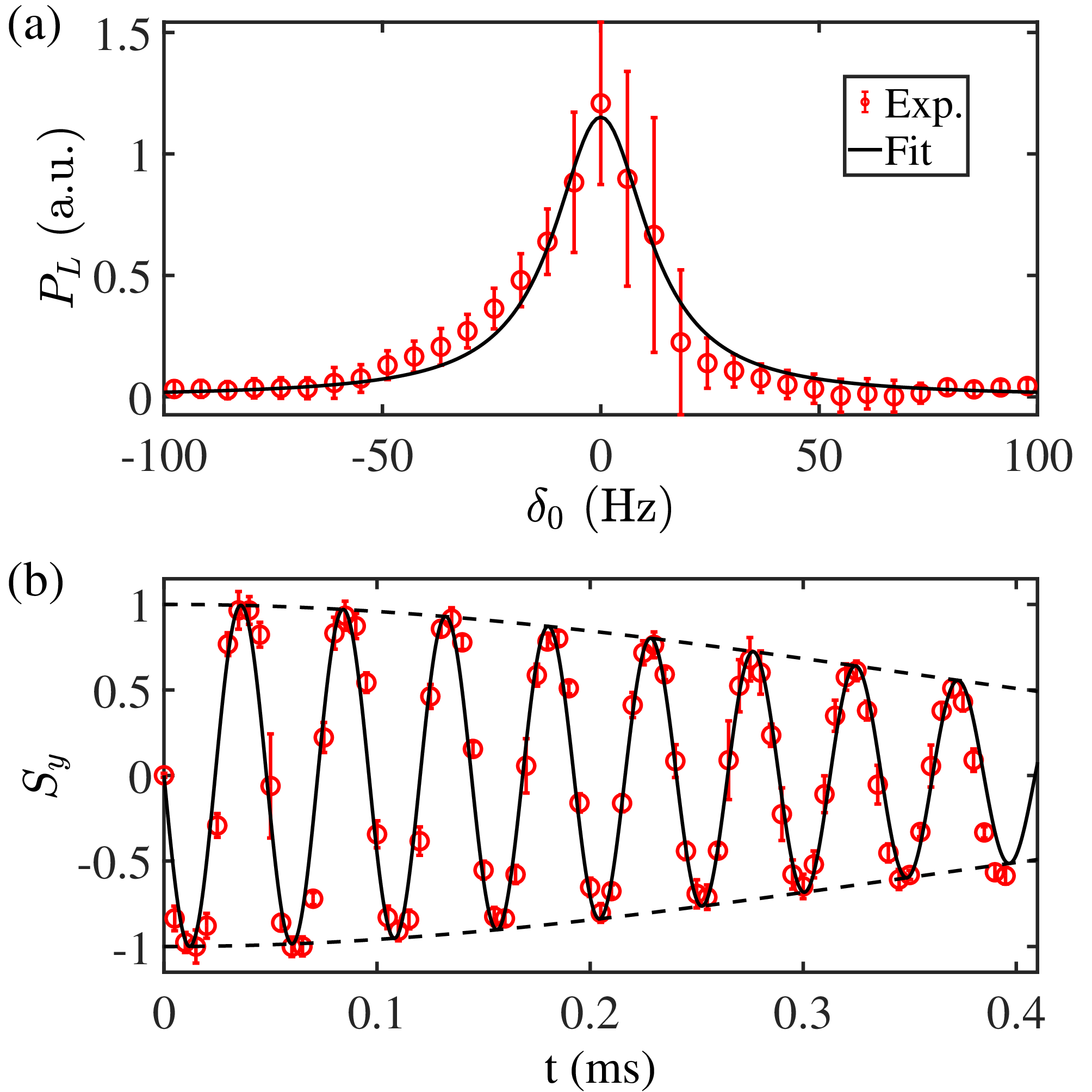}
	\caption{(a) Spectrum of NMR experiments (red circles with error bars) and fitting result (black solid line). The fitting function is assumed to be Lorentzian $P_L(\delta_0) = [\Gamma/(2\pi)] / [ \delta_0^2 + (\Gamma/2)^2]$ with a full width at half maximum (FWHM) $\Gamma = 0.026$~KHz. (b) Rabi oscillation of $S_y$ under the RF driving pulse with a nominal Rabi frequency $\Omega_R= (2\pi \times )21$~KHz (red circles with error bars). The black solid line is the fitting result with a Gaussian profile, $S_y(b, t) = -\sin(\Omega t) \exp(-b^2t^2/2)$. The fitting parameters are $\Omega = (2\pi \times )20.79$~KHz and $b = 0.46$~KHz. Data points and error bars are averaged over 10 repeated measurements. }
	\label{f3}
\end{figure}

\section{\label{sec:Experiment}Experimental realization in an NMR system}
To implement an off-resonance $\pi$-rotation and highlight the noise-resistance advantages of the NREP sequence, we carry out experiments in a liquid NMR system composed of a $C_2F_3I$ molecule ensemble. Along the $z$ axis, a strong bias magnetic field $B_0$ with a Larmor frequency $\omega_0 = (2\pi \times )39.638$ MHz lifts the degeneracy of each $^{19}F$ nuclear spin $I=1/2$, which is a two-level system spanned by the Zeeman states $|I = {1\over2}, m = -{1\over2}\rangle$ and $|I= { 1\over2}, m = {1\over2}\rangle$. We are able to distinguish each spin due to its large chemical shift difference, which is in the order of $(2\pi \times )1$~KHz. The weak coupling between these three spins $F_{1, 2, 3}$ is on the order of $(2\pi \times )0.1$~KHz~\cite{wang2025scientific}, which is negligibly small compared to the detuning $\Delta \sim (2\pi \times )21$~KHz and the driving-field Rabi frequency $\Omega \approx \Delta$. Therefore, we consider only the third nuclear spin $F_3$. The initial state of the spin is at thermal equilibrium, i.e., 
\begin{equation}
	\begin{aligned}
		\rho_T(0)={1\over 2}\sigma_0+{p\over 2}\sigma_{z},
	\end{aligned}
    \label{eq5}
\end{equation}
where $\sigma_0$ represents a $2\times 2$ identity matrix, and $p$ is the probability distribution satisfying  $|p| \approx \omega_0/(2k_BT_0) \sim 10^{-5}$ with the Boltzmann constant $k_B$ and the temperature of the sample $T_0 = 308$~K, i.e., $35^{\circ}{\text C}$. In the experiments, we measure the transversal nuclear magnetization ${S}_\perp = \sqrt{{S}_x^2 + {S}_y^2}$ which is proportional to the transversal average of the spin $F_3$.

We extract the noise strength of $\delta_0$ in the NMR experiments by fitting the Fourier transform of the free induction decay (FID) signal after rotating the spin in the $xy$ plane. As shown in Fig.~\ref{f3}(a), the spectrum is well described by a Lorentzian fitting function $P_L(\delta_0)$ with $\Gamma = 0.026$~KHz. Similarly, the noise strength of $\delta_1$ is extracted by performing a Rabi oscillation of $S_y$ under the resonant RF pulse with the Rabi frequency $\Omega_R= (2\pi \times )21$ KHz. Owing to power fluctuations of the RF pulse, the oscillation amplitude of $S_{y}$ rapidly decays on a short timescale of milliseconds, as shown in Fig.~\ref{f3}(b). By modeling the transverse noise $\delta_1$ with an oscillatory Gaussian function $P_G(b,t)$, we estimate that the noise strength of $\delta_1$ is $b = 0.46$~KHz. Apparently, the noise $\delta_0$ is typically as small as 5\% of $\delta_1$ and both are much smaller than $\Omega_R$, i.e., $\beta \ll \epsilon \ll 1$. Therefore, we may consider only the influence of the noise $\delta_1$ when we fit the data of our experiments with a simpler theoretical formula. 

\iffalse
Theoretically, we further compare the final state fidelity obtained via the DP and NREP sequences for spins initially along the $x$, $y$ and $z$ axis. It is expected that $F$ after the NREP sequence exceeds that after the DP sequence for all three representative initial spin states.
\fi

By fixing the detuning $\Delta=(2\pi \times )21$~KHz and the single-pulse duration $T=16.84$ $\mu$s, we implement both the DP and the NREP sequence for various Rabi frequencies $\Omega$, which deviate from the nominal one $\Omega_R$ by a Rabi frequency deviation $\delta_R = \Omega-\Omega_R$ with $\Omega_R = \Delta$. In this manner, we investigate the robustness of the DP and the NREP sequence by changing the deviation $\delta_R$ gradually from zero. We measure the transversal magnetization signal $S_{\perp}$ in the $xy$ plane after the DP or the NREP sequence for two initial spin configurations. One is that all three spins are aligned along the $+z$ direction (the thermal state Eq.~(\ref{eq5})), and the other is that spin $F_3$ is along the $+z$ direction and spins $F_{1,2}$ are roughly zero in average. The latter spin configuration is achieved by applying a $\pi/2$-pulse to spin $F_1$ and $-\pi/2$-pulse to spin $F_2$ about $y$-axis starting from the thermal state and waiting for a delay of about $2$~s. Spins $F_1$ and $F_2$ become completely dephased due to the longitudinal noise $\delta_0$, but do not yet undergo significant relaxation since the nuclear relaxation time $T_1 \sim 10$~s and the inhomogeneous dephasing time $\Gamma^{-1} \sim 30$~ms~\cite{wang2025scientific}. Although the scalar couplings among three spins are weak, the latter spin configuration may further mitigate their influence, compared to the former spin configuration. Once the transversal magnetization is measured, the final state fidelity is calculated according to the formula $F = \left(1+\sqrt{1-S_\perp^2}\,\right)/2$.

\begin{figure*}[tbhp]
	\centering
	\includegraphics[width=7in]{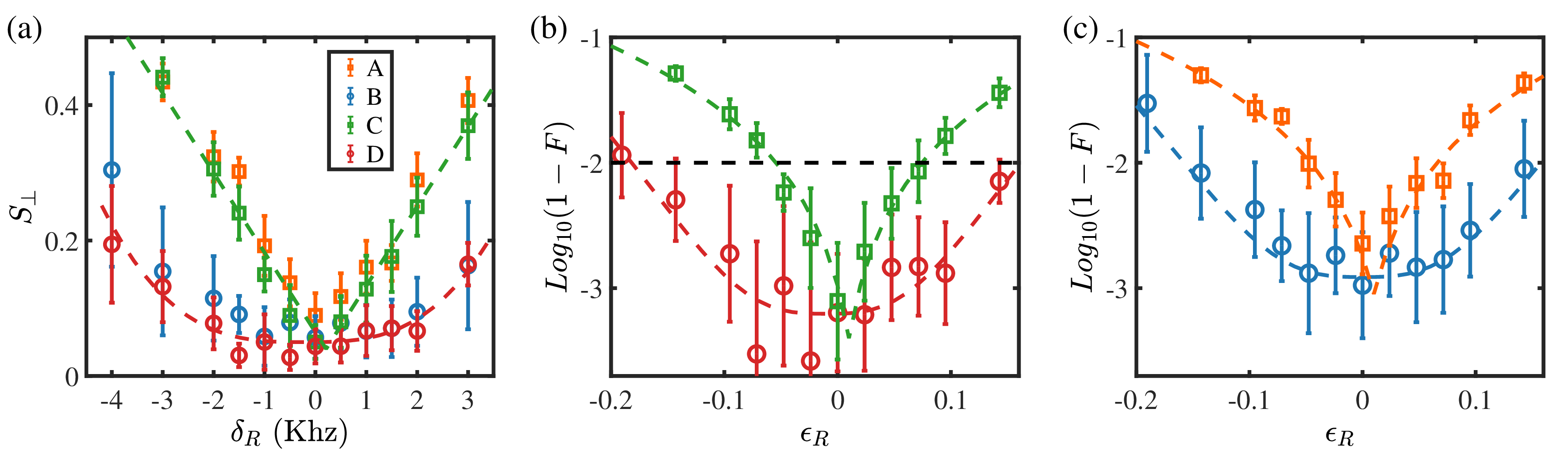}
	\caption{(a) Dependence of $S_{\perp}$ after either the NREP (circles with error bars) or the DP sequence (squares with error bars) on the Rabi frequency deviation $\delta_R$ with $\delta_R=\Omega-\Omega_R$ and $\Omega_R= (2\pi \times )21$~KHz. Spins $F_1$ and $F_2$ are initially both along the $+z$ direction in experiments A and B, while they are roughly zero in experiments C and D. The red and green dashed lines are the fitting results for the NREP and the DP sequence, respectively. Final state fidelity in experiments (b) C and D, and (c) A and B, where the relative Rabi frequency deviation is defined as $\epsilon_R=\delta_R/\Omega_R$. The dashed curves in (b) and (c) are the corresponding fitting results. Experimental data points and error bars (standard deviation) are obtained from 100 measurements. Obviously, the highest fidelity of the NREP and the DP sequence both are near 99.9\%,  while the noise-tolerance range of NREP is about 3 times as wide as that of DP as characterized by the horizontal black-dashed line in (b).}
	\label{f4} 
\end{figure*}

We plot the experimental data in Fig.~\ref{f4}(a). For the DP or the NREP sequence, both initial spin configurations show close results, implying that the scalar couplings among spins are weak and have little effect on the performance of the pulse sequence. We fit the experimental data with a theoretical formula, $S_\perp \sim |\delta_R|^3$ (detailed in the SM~\cite{sm}) for the NREP sequence, and $S_\perp \sim |\delta_R|$ for the DP sequence. The fitting parameters agree well with theoretical predictions. 

With the transversal spin $S_\perp$, we calculate the final state fidelity $F$ and plot it in Fig.~\ref{f4}(b) and (c) for the two spin configurations, respectively. The highest fidelity reaches $0.9997 \pm 0.0006$ after the NREP sequence and $0.9992\pm 0.0008$ after the DP sequence as shown in Fig.~\ref{f4}(b). This indicates that both the NREP and the DP sequences are capable of realizing an off-resonance $\pi$ rotation with a high precision. Importantly, the noise-tolerance range of the NREP sequence is about 3 times as wide as that of the DP, which is also consistent with the theoretical prediction. These findings clearly manifest the superior power of the NREP sequence in noise suppression while realizing the off-resonance $\pi$ rotation.

\section{\label{sec:Conclusions and discussions}Conclusions and discussions}
Under the off-resonance condition, we propose two pulse sequences to achieve a high-precision $\pi$-rotation of a spin, the DP and the NREP sequences. The NREP sequence simultaneously suppresses to the fourth order the transversal noise from the driving RF field and the longitudinal noise from stray magnetic fields, while the DP suppresses them to the second order. As a proof-of-concept, we experimentally demonstrate in a liquid NMR system the off-resonance $\pi$-rotation under the NREP sequence. The highest fidelity reaches $0.9997 \pm 0.0006$ for the polarized thermal initial spin state. The noise-resistance region (above 0.99 fidelity) is as wide as about $20\%$ of the driving field strength.

The off-resonance $\pi$-rotation provides a useful method to solve the crosstalk problem in a two-qubit system. For two qubits (spins), when we rotate the target qubit $Q_1$ with a resonant driving field, the idle qubit $Q_2$ with a detuning  $\Delta$ is also affected slightly, due to the control crosstalk~\cite{RevModPhys.76.1037}.  With the NREP sequence, one may adjust the driving field frequency to match the resonance of $Q_2$. After the NREP sequence, the target qubit $Q_1$ performs an accurate $\pi$-rotation while the idle qubit $Q_2$ is effectively left untouched. Furthermore, by adjusting the driving frequency to be exactly the middle frequency between $Q_1$ and $Q_2$, a two-qubit $\pi$-rotation can be simultaneously achieved under the NREP sequence. In this manner, one may avoid the errors caused by a hard pulse-driving~\cite{Cummins_2000} or complex selective pulse sequences~\cite{PhysRevA.107.032618}. The NREP sequence also provides a new approach to realize a high-precision $\pi$-rotation in quantum-dots where the longitudinal and transversal noises coexist~\cite{articleWang,articleFoletti}.

\begin{acknowledgments}
	The work is supported by the National Natural Science Foundation of China (NSFC) under Grant Nos. 92576103 and 12274331, the Quantum Science and Technology - National Science and Technology Major Project (Grant No. 2021ZD0302100), and the HZNU scientific research and innovation team project (TD2025003). The numerical calculations in the paper have been partially done on the supercomputing system in the Supercomputing Center of Wuhan University.
\end{acknowledgments}

\section*{Data Availability Statement}
The data that support the findings of this article are not publicly available. The data are available from the authors upon reasonable request.
%\begin{thebibliography}{71}%
%\makeatletter
%\bibliographystyle{apsrev4-2}
\bibliography{opv}

\end{document}